\documentclass{aa}  
\usepackage{natbib}
\usepackage{graphicx}

\usepackage{txfonts}
\usepackage{xcolor}

\defcitealias{delaReza25}{DLR25}

\begin{document}

   \title{Lithium abundance and  stellar rotation  in the \\ Galactic  halo and thick disc} 
    \subtitle{Contribution from low-mass giant field stars}

   \author{R. de la Reza,
          \inst{1}
          F. Llorente de Andrés\inst{2}
          \and
          E. J. Alfaro\inst{3}
           \and
        C. Chavero\inst{4,5}
          }

   \institute{ Observatório Nacional, Rua General José Cristino 77, 28921-400 São Cristovão, Rio de Janeiro, RJ, Brazil\\
              \email{ramirodelareza@yahoo.com}  
        \and
        Centro de Astrobiología (CAB), CSIC-INTA, Camino Bajo del Castillo s/n, Campus ESAC, 28692, Villanueva de la Cañada, Madrid, Spain 
        \and
             Instituto de Astrofísica de Andalucía, CSIC, Glorieta de la Astronomía, s/n, Granada 18008, Spain 
        \and
              Observatorio Astron\'omico de C\'ordoba, Universidad Nacional de C\'ordoba, Laprida 854, 5000 C\'ordoba\\  
             \email{carolina.chavero@unc.edu.ar}
        \and
       Consejo Nacional de Investigaciones Científicas y Técnicas (CONICET), Godoy Cruz 2290, Ciudad Autónoma de Buenos Aires, Argentina
             }

   \date{Received  accepted }

 
\abstract 
{
The stellar evolution of lithium-rich (Li-rich) giant stars at very low metallicities remains largely unexplored to date. Using mainly two recent large LAMOST catalogues of field, low-mass giant stars (both Li-rich and Li-poor) with metallicities ranging from  $-4.0$ up to $-1.0$, we studied some of the conditions for Li enrichment and the general distribution of stellar rotations in the Galactic halo and thick disc. Due to the scarcity of stars  with [Fe/H] $< -3.0$, only three  Li-rich red giant branch (RGB) stars are known in this regime. The full observational appearance of all giants, across the horizontal branch  (HB) and asymptotic giant branch (AGB) advanced stages of evolution (with Li abundances  up to $6.15$ dex)  have been detected for metallicities $> -2.5$.
Among these stars, we detected the presence of IR excesses that 
are considered to be indicative of giant stars losing mass,
showing a recent episodic Li-enrichment process related to the Cameron-Fowler mechanism for the formation of new \( ^7\text{Li}\).
Because stars presenting IR excesses are distributed across the majority of metallicity values, we suggest this
mechanism is at work throughout an important part of the evolutionary history of the Galaxy.  Based on these IR excesses, we identified three Li thresholds: $\sim$ 1.5~dex for RGB stars, $\sim 0.5$~dex for HB stars, and $\sim -0.5$~dex for AGB stars, thereby establishing a new criterion to characterise Li-rich giants in the halo and thick disc. We carried out a first extensive study of stellar rotations in metal-poor giant stars, revealing the following results:
a)  a plateau appears   for velocities greater than $40\,\mathrm{km\,s^{-1}}$ extending up to near $90\,\mathrm{km\,s^{-1}}$, with Li abundances ranging from  1.02  to 1.82 dex;
b) among Li-rich giant stars with  $v \sin i > 40\,\mathrm{km\,s^{-1}}$, a clear trend toward increasing rotation is observed up to near $90\,\mathrm{km\,s^{-1}}$, as metallicities decrease from $-1.0$ to $-2.5$, it is observed; c) the presence of RGB and HB Li-rich giant stars with  rotations up to $90\,\mathrm{km\,s^{-1}}$ suggests that internal stellar models must account for extended
\( ^3\text{He} \) reservoir lifetimes
as a source of \( ^7\text{Li}\) considering these velocities. The velocity around $40\,\mathrm{km\,s^{-1}}$ appears to be a new critical value that merits further investigation.}

   
   
   
   


 \maketitle
   
\section{Introduction}

The fragile light element lithium (Li), with its cosmological and stellar significance in the evolution of the Universe, has been extensively studied throughout the literature. With respect to cosmological constraints, the main issue is related to the persistent discrepancy between the primordial Li abundance and the values derived from metal-poor halo stars. Recent results indicate a primordial Li abundance of $2.69 \pm 0.02$ \citep{Yeh21} and a stellar Li abundance of $2.16 \pm 0.07$ \citep{Matas-P21}.

In general studies of Galactic Li, significant differences arise depending on the metallicity regime under consideration,  with dwarf stars at very low halo metallicities having  primarily been associated with studies of the well-known Spite plateau.
With respect to giant stars and in relation to the cosmological Li problem, \citet{Mucciarelli2022} found a thin Li plateau at a Li abundance of around 1.0 dex. This plateau, with metallicities between approximately -4.0 and -1.0 formed by lower red giant branch (RGB) stars, differs from that of dwarf stars.
However, it points to the same Li value as in dwarfs once dredge-up is accounted for. Presently, we find ourselves in a unique situation in which only a few halo and thick-disc Li-rich giant stars have been discovered, making the study of such stars at very low metallicities a relatively new and open field  for research.

Since 2008, several observers have identified new Li-rich field giant stars at very low metallicities, generally referred to as  low-mass stars ($M < 2 M_{\odot}$). These detections include: \citet{Roederer08, Martell13, Ruchti11, Roederer14, Li_Cat18, Sitnova23, Susmitha24, Kowkabany24}, and \citet{Mu24}. As noted by these authors, the Li abundances of these giant stars range from $1.6$ to $4.8$, with $\mathrm{[Fe/H]}$ values from $-4.0$ to $-1.4$, and $\log g$ values from $0.6$ to $3.2$. A comprehensive catalogue of these stars was compiled by \citet{Mu24}, listing approximately 30 Li-rich field giant stars with these parameters.

A key physical aspect related to Li-rich giant stars at very low metallicities concerns the mechanism responsible for the production of fresh \( ^7\text{Li}\). The  extent to which these processes are similar to those acting in the higher-metallicity regime ($\mathrm{[Fe/H]} > -1.0$), such as those in the Galactic thin-disc, is a matter of debate. Two main scenarios have been proposed thus far: (1) an internal stellar mechanism based on the well-known Cameron-Fowler (CF) process or (2) an external mechanism involving mass transfer from a more massive companion in a binary system. There is no clear consensus on which scenario is the most plausible.
A recent study by \citet{Sayeed2025}
examined the different types of binaries that contribute to the Li
enhancement. Based on GALAH observations, their main results indicate that
binary stars can produce Li-rich giant stars with A(Li) values ranging from 1.5
dex to 2.2 dex; however, to attain these values, these stars must be enriched in
some way to achieve Li abundance values between 2.5 dex and 3.2 dex during
their main sequence stage. Furthermore, to obtain Li-rich or very Li-rich giant
stars with Li values greater than 2.2 dex, these stars require additional
mechanisms.

Other important parameters discussed in the literature include the scarcity of data on stellar rotation and activity in the very low-metallicity regime. Nevertheless, some recent exceptional discoveries of Li-rich giants provide input for further research in this emerging field. For example, the Li-rich giant with the lowest metallicity discovered so far, HE~0057-5959 \citep{Mu24}, has $\mathrm{[Fe/H]} = -3.95$, a Li abundance of $2.05$ dex and a $\log g$ value of $3.05$. Another remarkable case is 2MASS J05241392-0336543, which has a Li abundance of $6.15$ dex \citep{Kowkabany24} and $\mathrm{[Fe/H]} = -2.43$, making it the most Li-rich giant star detected to date.

In this work, we investigate the evolution of Li-rich giant stars in the halo and thick-disc Galactic regions. Our study is based on two datasets of low-mass giant stars with metallicities between $-4.0$ and $-1.0$: (a) a sample of 1703 Li-rich giant stars with Li abundances between $1.5$ and $6.15$ dex, primarily obtained from the LAMOST DR9 catalogue \citep{Ding24}, with a few additional stars from the literature; and (b) a dataset of 2468 Li-rich and Li-poor giant stars, resulting from merging the LAMA LAMOST catalogue \citep{Li24} for rotational velocity ($v \sin i$) values with the LAMOST DR9 catalogue \citep{Ding24} for Li abundances. We conducted a preliminary analysis of the membership of the selected stars in different Galactic subsystems (halo, thick disc, and thin disc) based on \textit{Gaia} DR3 astrometric data \citep{GaiaDR3}, metallicity, and the alpha-element-to-iron ratio.

This paper is structured as follows:  
Section~2 analyses and compares the values of the physical variables
used, which were taken from the catalogues compiled by \citet{Ding24} and \citet{Li24}.
It also evaluates and discusses the uncertainties
associated with the variables involved in our study and examines the
membership of these objects in the different Galactic subsystems.
Section 3 discusses the
evolutionary status of Li-rich giant stars. Section 4 provides an overview of the
Cameron-Fowler model. Section 5 presents a general discussion on stellar
rotation. Finally, Sect. 6 offers our conclusions and a summary.

\section{ Sample analysis: Bias, errors, and kinematic classification}

\subsection{Bias and errors} 

Our analysis is based mainly on the LAMOST DR9 catalogue
published by \citet{Ding24}. This survey contains
information on the parameters [Fe/H], $\log(g)$, $T_{\text{eff}}$, RV, and A(Li), and it is available in Vizier under
the reference \texttt{J/ApJS/271/58/table1}. Only the Li
abundance has associated parameters that quantify both the
precision and the uncertainty of the values, in comparison
with the set of standards adopted and other previous
catalogues derived from LAMOST spectra. We adopted
the variable \texttt{e\_DelA(Li)} as a reasonable estimate of
the uncertainty in the Li abundance.
 Its median value calculated for our total number of 3537 stars is 0.1 with a SD of 0.3.  
The projected stellar rotational velocities, $V \sin i$, were
taken from \citet{Li24}. This catalogue is also based on the
LAMOST medium-resolution spectra (MRS), as in \citet{Ding24}. Both catalogues provide [Fe/H], $T_{\text{eff}}$, $\log(g)$, and RV values, derived through different
methodologies and, crucially, using different sets of
calibration stars. \citet{Li24} also provides an individual
uncertainty estimate of $V \sin i$ for each object.
The availability of two catalogues with stellar parameters
derived from the same set of spectra allows us to estimate
statistical indicators that quantify the uncertainty of the
parameters used in this study; specifically, those from \citet{Ding24}. Cross-matching both catalogues for our target sample yields
2465 stars in common. The statistical estimators used are
defined below, with the resulting values given in Table 1.\ \  
\begin{itemize}
\item $\Delta$ Sys – median of the difference
between the value in \citet{Ding24} and that in \citet{Li24};
\item MAD – median of the absolute value of the
difference, corrected for the estimated bias;
\item $\sigma_{\mathrm{crit}}$ – dispersion
calculated as $1.4826 \times \mathrm{MAD}$, assuming a
Gaussian distribution;
\item \% outliers – fraction of stars whose absolute
bias-corrected difference exceeds $3\,\sigma_{\mathrm{crit}}$;
\item $\sigma_{\mathrm{internal}}$ – median of
the individual uncertainty reported by \citet{Li24}.
\end{itemize}

\begin{table}[ht]
\centering
\caption{Comparison of stellar parameters from \citet{Ding24} and \citet{Li24} for 2465 common stars.}
\label{tab:tab1}
\small
\setlength{\tabcolsep}{4pt} %
\begin{tabular}{lrrrrcr}
\hline\hline
Parameter & $N$ & $\Delta$ Sys & MAD & $\sigma_{\mathrm{crit}}$ & \% outliers & $\sigma_{\mathrm{int}}$ \\
\hline
{[Fe/H]} (dex) & 2465 & -0.16 & 0.13 & 0.19 & 4 & 0.08 \\
$T_{\mathrm{eff}}$ (K) & 2465 & -20 & 102 & 151 & 2 & 40 \\
$\log(g)$ (dex) & 2465 & 0.00 & 0.22 & 0.33 & 3 & 0.14 \\
RV (km\,s$^{-1}$) & 2465 & -4.6 & 1.00 & 1.48 & 4 & 0.18 \\
\hline
\end{tabular}
\end{table}

These results show that discrepancies arising
from the use of different pipelines and calibration standards
introduce greater uncertainties than individual random
errors associated with the spectra quality and data reduction
methodology. The largest systematic offsets are found for
metallicity and radial velocity.
The projected rotational velocity, $V\sin i$, comes directly from
\citet{Li24}, along with an individual uncertainty
estimate for each object. The median of these uncertainties is
$1.76$ km\,s$^{-1}$.
For the kinematic analysis, we used stellar distances, proper
motions, and radial velocities. Distances were taken from \citet{Bailer21}, proper motions from \textit{Gaia} 
DR3, and RVs from \citet{Ding24}. Distances and proper
motions have individual uncertainty estimates. For all stars,
we adopted an RV uncertainty of 7 km\,s$^{-1}$, based on
the bias between the two catalogues and a representative
value of the combined systematic and random errors (see
Table~\ref{tab:tab1}). Assuming quadratic error
propagation, we derived the distribution of relative velocity
errors, finding that 90\% of the stars in our sample have a
relative velocity error below 16\%.

\subsection{Sample distribution in the galactic phase space}

We distinguished between three different evolutionary stages: RGB
stars, horizontal branch (HB) stars, and asymptotic giant branch (AGB) stars. This
classification is based on surface gravity ($\log g$), following the methodology of \citet[hereafter DLR25]{delaReza25} 
and the asteroseismic study by \citet{Yan21}. Specifically, we applied:\ 
\begin{itemize}
\item RGB stars: $2.8 \leq \log g \leq 3.5;$
\item HB stars: $2.0 \leq \log g < 2.8;$
\item AGB stars: $0.3 \leq \log g < 2.0.$
\end{itemize}

Our main objective was to determine the distribution of the three classes of
giant stars used in our study across the different Galactic subsystems: thin disc,
thick disc, and halo. The ranges
of alpha-to-iron abundance ratio, [$\alpha$/Fe], and metallicity, [Fe/H], further
constrained the classification. 
Nevertheless, it must be noted that this classification method is only an approximation. For instance, it cannot unambiguously distinguish whether a star is an AGB or an upper-RGB star.
We obtained the variables (X, Y, Z, U, V, W), referred to the Sun, for the catalogue by \citet[LAMOST-DR9]{Ding24}, which contains the chemical abundances A(Li) and [Fe/H], forming the basis of our study. Radial velocities were taken from the same catalogue, while the equatorial ICRS coordinates and proper motions originate from \textit{Gaia} DR3 \citep{GaiaDR3}. Distances were taken from \citet{Bailer21}, based on \textit{Gaia }EDR3 parallaxes \citep{GaiaEDR3}. The transformation to a Galactocentric reference frame was carried out assuming $R_0 = 8.2$ kpc, a solar rotation velocity of $V_{\mathrm{LSR}} = 232$ km \,s$^{-1}$ \citep{McMi17} and a peculiar solar motion of $(U_0, V_0, W_0) = (5.0, 17.5, 6.46)$ km \,s$^{-1}$, determined from the kinematics of cool giant stars \citep{Ayari24}. With these values, we computed the Galactocentric spatial coordinates $(X_{\mathrm{gc}}, Y_{\mathrm{gc}}, Z_{\mathrm{gc}})$ and the cylindrical velocity components with respect to the Galactic centre $(V_R, V_\phi, V_Z)$.

\begin{figure}
    \centering
\includegraphics[width=0.48\textwidth]{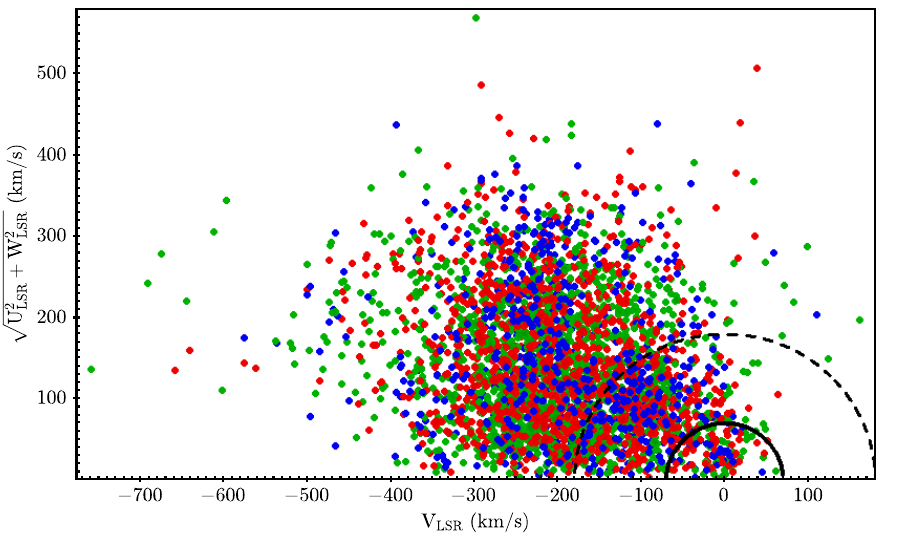}
    \caption{Diagram of Toomre for the  stellar samples used in this work. Green indicates AGB stars, red represents HB stars, and blue corresponds to RGB stars. The two semicircles indicate, as a first approximation, the boundaries between the thin and thick disc (70 km\,s$^{-1}$), and between the thick disc and the halo (180 km\,s$^{-1}$). }
    \label{fig:Toomre}
\end{figure}

Using this information, we present a Toomre diagram (see Fig.~\ref{fig:Toomre}), where the classes defined by intervals of $\log g$ appear in green (AGB), red (HB), and blue (RGB). Semicircles indicating approximate boundaries between the three Galactic components present in the sample are overplotted. The boundary between the thin and thick disc is set at 70 km\,s$^{-1}$, while the boundary between thick disc and halo is set at 180 km\,s$^{-1}$ \citep{Mardini22}. 

\begin{figure}
    \centering
    \includegraphics[width=\columnwidth]{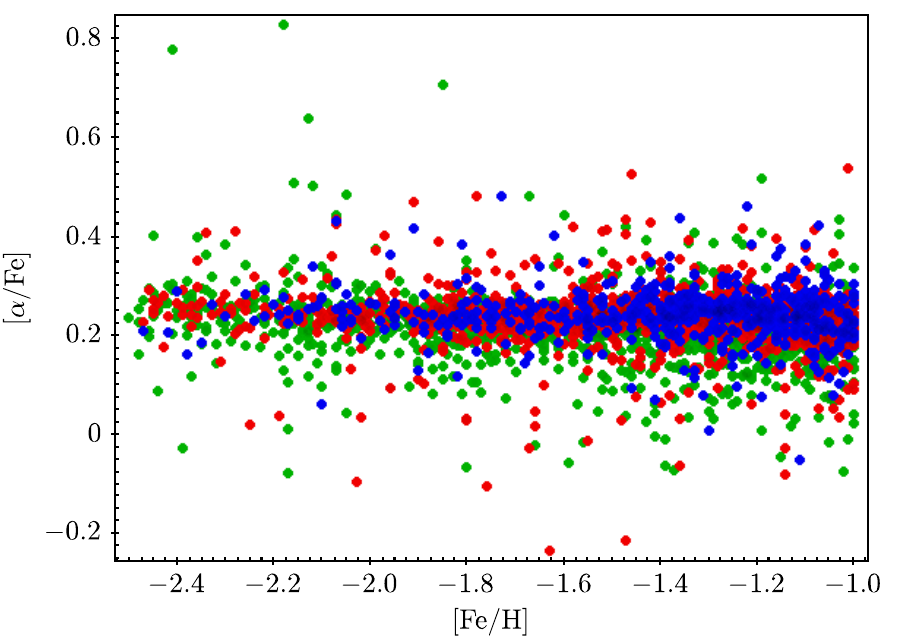}
    \caption{[$\alpha$/Fe] versus [Fe/H]. Most of the stars are concentrated around [$\alpha$/Fe] $=$ 0.23 (median) with a $\sigma =$ 0.07. Green indicates AGB stars, red represents HB stars, and blue corresponds to RGB stars.}
    \label{fig:alfa_Fe}
\end{figure}

Here, [$\alpha$/Fe] as a function of [Fe/H] is shown in Fig.~\ref{fig:alfa_Fe}. The [$\alpha$/Fe] values are taken from \citet{Zhang24}, while [Fe/H] is from \citet{Ding24}. As expected for halo and thick-disc stars, [$\alpha$/Fe] is generally positive. It is interesting to note the presence of some negative values, indicating a deficiency in $\alpha$-elements across the entire  low metallicity range. The most plausible interpretation for these $\alpha$-deficient giant stars is contamination by $\alpha$-poor supernovae of Type Ia and pure core-collapse events \citep{Jeena2024}.

\begin{figure}
    \centering
    \includegraphics[width=\columnwidth]{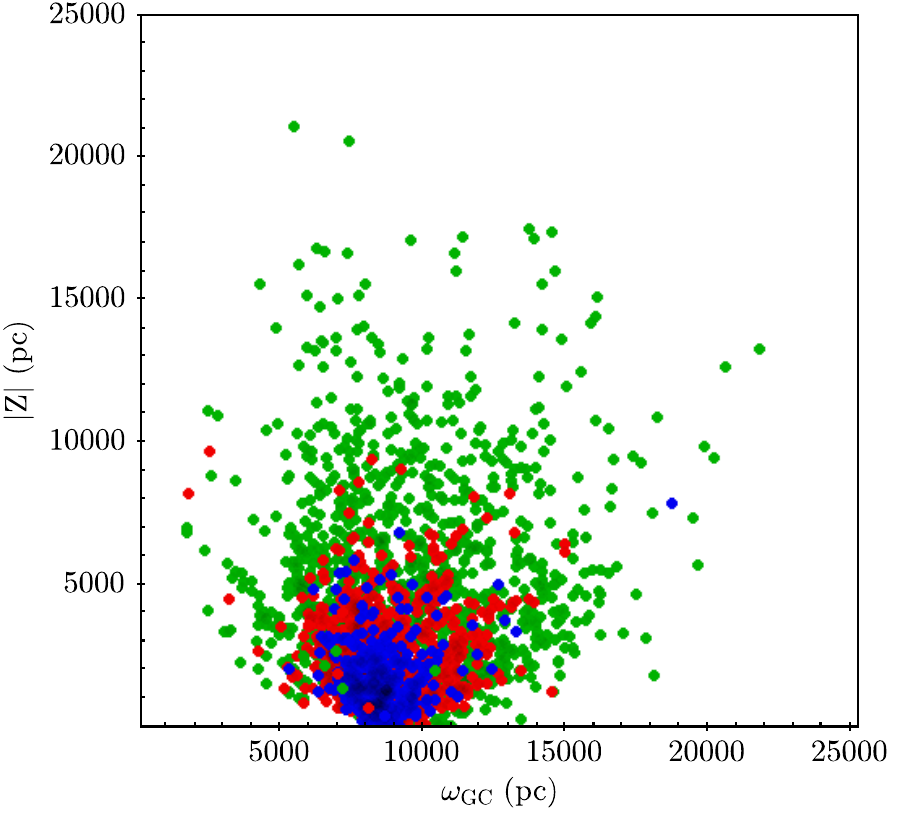}
    \caption{Spatial distribution of the giant stars used in this work. Here, $\omega_{\mathrm{GC}}$ is the Galactocentric radius projected onto the Galactic plane and $|Z|$ is the absolute distance from the mid-plane. The radial extent of each stellar class increases as $\log g$ decreases. Colours and symbols are the same as in Fig.~\ref{fig:Toomre}.}
    \label{fig:Z_Omega}
\end{figure}

According to these data, we obtained the following classification: halo (74\%), thick disc (22\%), and thin disc (4\%). The stars associated with the thin-disc appear to belong to the so-called metal-poor thin-disc \citep[or Atari disc,][]{Mardini22}, with some having [$\alpha$/Fe] and [Fe/H] values suggestive of a primordial thin-disc origin. Figure~\ref{fig:Toomre} shows no significant difference between the three evolutionary stages of the giants. However, the spatial distribution of these objects in Galactocentric cylindrical coordinates $(|Z|, \omega_{\mathrm{GC}})$ (see Fig.~\ref{fig:Z_Omega}) reveals a well-defined pattern in which the spatial extent of each class increases with decreasing $\log g$, as expected since the luminosity of giants increases with decreasing $\log g$ for the selected sample. In summary, there appears to be no underlying metallicity pattern in either the Toomre diagram or the spatial distribution of the objects in our samples.

\section{The stellar evolutionary stages of Li-rich giant stars}

The exploration of the Galactic halo provides insight into the subsequent generations of the first stars in the Galaxy. These very old stars are considered fossil records of early chemical enrichment. Although the primordial light element  Li is a key component of these stars, investigations of carbon, the first heavy element synthesised after the Big Bang, have also driven the study of early chemical enrichment. However, these very metal-poor stars are challenging to detect because of their rarity, representing only about $0.1\%$ of all stars in the Galaxy \citep[][and references therein]{Yao24}.

The extreme scarcity of stars in the metallicity range $-4.0 \leq \mathrm{[Fe/H]} \leq -3.0$ is evident in the Galactic maps by \citet[see their Fig.~3]{Chiti21}. In recent years, searches based on spectroscopy and photometry have identified only a modest number of stars within the metallicity range $\sim -4.0$ to $\sim -2.0$. It is only with large-scale surveys such as \textit{Gaia} (photometry) and LAMOST and APOGEE (metallicity) that these searches were significantly advanced. Thanks to these large datasets, extensive catalogues of halo  RGB stars \citep{Viswanathan24} and very metal-poor stars \citep{Yao24} have been compiled.

\begin{figure}[ht]
\centering
\includegraphics[width=\columnwidth]{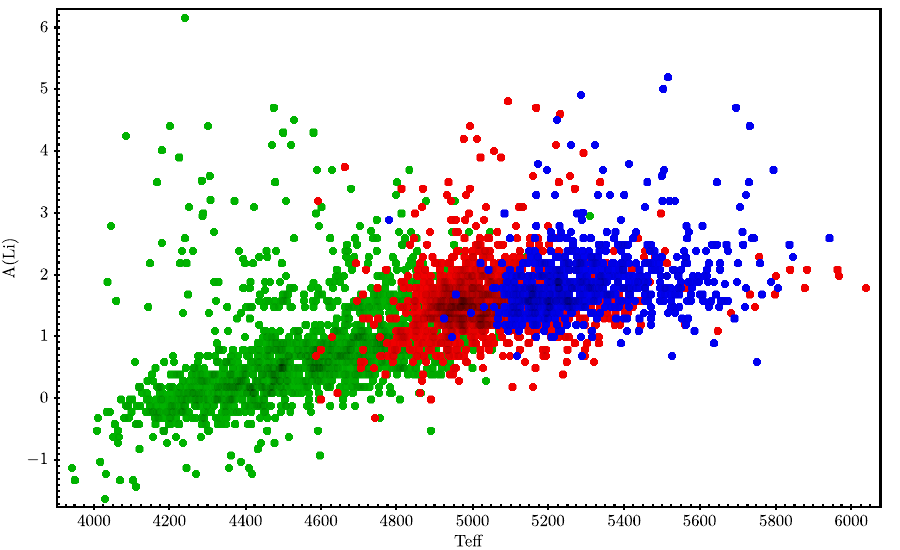}
\caption{Li abundance as a function of $T_{\text{eff}}$. See the main text for details.}
\label{fig:Li_Teff}
\end{figure}

Figure~\ref{fig:Li_Teff}  presents the distribution of all stellar data with $\mathrm{[Fe/H]} \leq -1.0$  used in this work, 
showing the Li abundance as a function of $T_{\text{eff}}$ for the catalogue of \citet{Ding24}. 
AGB stars (green), HB stars (red), and RGB stars (blue) exhibit a clear relationship to $T_{\text{eff}}$
for stars cooler than 6000 K. Notably, this dataset aligns well with our sample of halo and thick-disc stars.

The detection of Li-rich giant stars in the extreme metallicity range $-4.0 \leq \mathrm{[Fe/H]} \leq -3.0$ is even more challenging, as these objects only account for $\sim 1.0\%$ of normal Li-poor giants  in the thin disc (see \citetalias{delaReza25}). The most important catalogues of giant stars are represented by the LAMOST and GALAH catalogues, which have included Li abundances; in
LAMOST, other parameters such as the projected stellar rotation velocities are also provided. This new data  has enormously changed
our panorama on field low-mass giant stars.

Among the LAMOST catalogues discussed in this work, the first important resource is that work of Gao et al. (2019), based on low-resolution spectra.
A new version later appeared,  based on medium-resolution spectra, published in \cite{Gao21}. Both studies used a
minimum Li abundance of 1.5 dex to characterise a Li-rich giant star.
In 2024, two additional catalogues based on medium-resolution spectra were released:
the LAMA LAMOST catalogue  \citep{Li24},  which introduced
the stellar rotation values, $v \sin i$, and the LAMOST DR9 catalogue \citep{Ding24},  which significantly expanded the number of targets. These  last two
catalogues were used in this work. 
Here, we present a comparison between the data from \cite{Ding24} and \cite{Gao21}  to highlight important
differences  related to  physical features reported in Gao’s catalogues.

\begin{figure}
    \centering
    \includegraphics[width=\columnwidth]{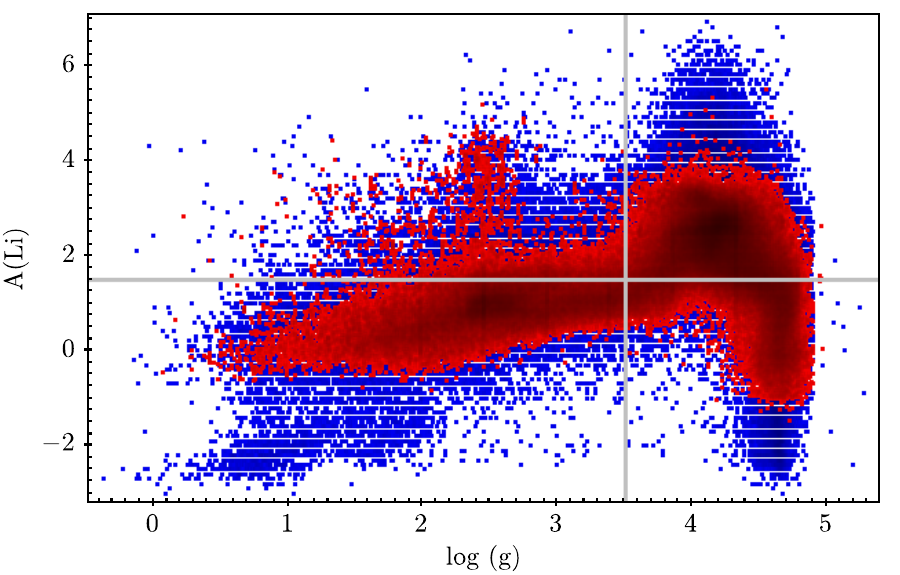}
    \includegraphics[width=\columnwidth]{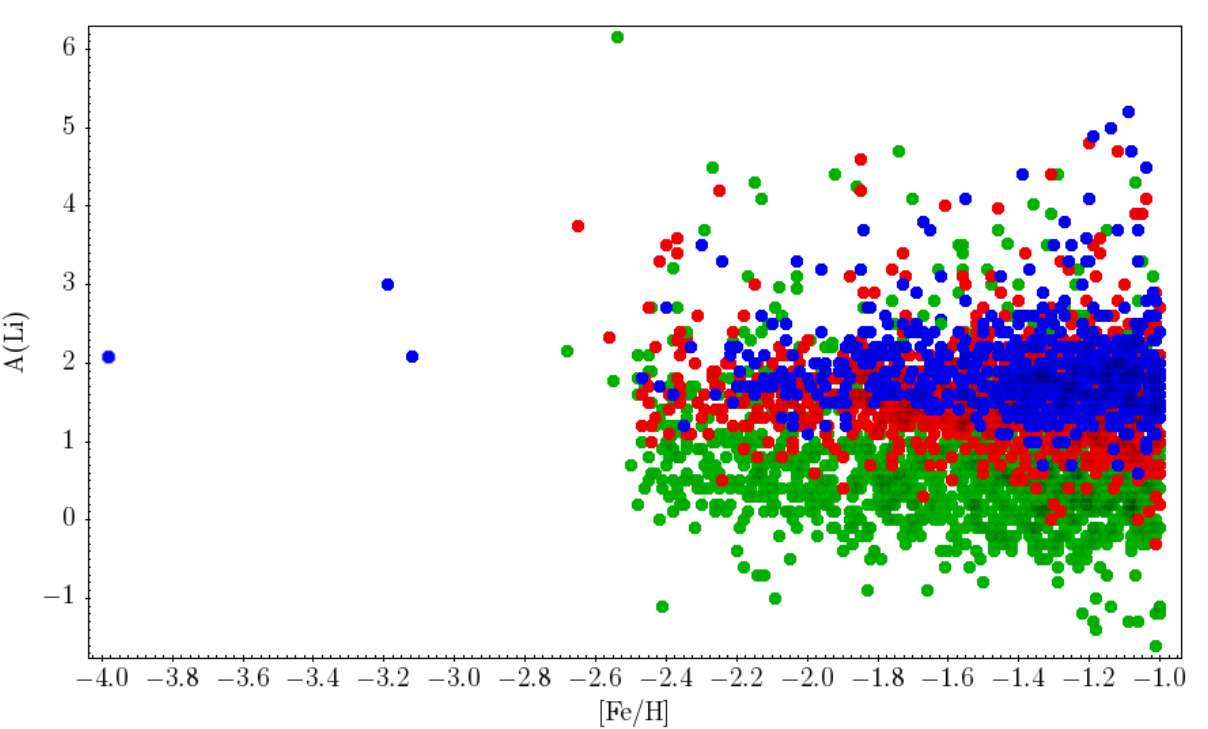}

    \caption{
        Li abundance trends in evolved stars.
        Panel a (top): General stellar distributions of the catalogues of \citet[blue]{Ding24} and \citet[red]{Gao21}, showing Li abundances as a function of $\log g$. The horizontal line at A(Li)=1.5 serves only as a reference, while the vertical line marks the end of the stellar giant branch at $\log g = 3.5$.
        Panel b (bottom): Li abundance of Li-rich giant stars, RGB (blue), HB (red), and AGB (green), as a function of metallicity ([Fe/H]) in the range $-4.0$ to $-1.0$. While RGB stars are present throughout the metallicity range, HB and AGB stars appear only for [Fe/H] $> -2.5$.
    }
    \label{fig:Li_trends}
\end{figure}

First, concerning the Li abundance values, both catalogues reveal a similar
behaviour in the interval corresponding to those of Li-rich giant stars.
The only exception corresponds to very low Li abundances, namely, less than -1
(noting this exception is not shown in these figures). This means that the important Li
abundance values derived from both catalogues are similar.
In Fig.~\ref{fig:Li_trends}a, we show a representation of the Li abundance of field
giant stars as a function of log g for both catalogues: that of \cite{Gao21}
 in red and that of \cite{Ding24} in blue. In general, both
distributions present the same patterns, even considering the quite
different numbers of stars involved. However, concerning giant stars
that occupy the log g interval between $\log g$  values near 0.3 up to 3.5,
an important difference appears between $\log g$ at 2.8 and that at 3.5,
which coincides with our classified RGB stars. Specifically, in this
interval, Ding’s RGB stars are much more numerous than those
corresponding to Gao’s. In other words, RGB stars in Ding’s distribution
present many more Li-rich giant stars than those in Gao’s distribution.
The same  differences  appear for AGB stars defined
between $\log g$ values of 0.3 and 2.0. However, it is noteworthy that HB
stars defined around the $\log g $ value of 2.5 present similar distributions
in both catalogues. The consequence of the aforementioned difference in the
interval corresponding to RGB stars is as follows. \cite{Yan21} and
\citetalias{delaReza25}, which worked with Gao’s catalogues for higher metallicities
corresponding to those of the Galactic thin-disc, found that only RGB
stars present a maximum Li abundance near the value of $2.6$ dex. If we
go on to consider the previously mentioned large difference in RGB stars of both
catalogues, we can infer that working with Ding’s catalogue, this low Li
abundance will no longer be valid and RGB stars will attain very large Li
abundances similar to those of HB and AGB stars. In fact, this is what is
found here using Ding’s catalogue, as can be seen in Fig.~\ref{fig:Li_trends}b, where Li
abundance is shown as a function of metallicity for metallicities between
$-4.0$ and $-1.0$ corresponding to the Galactic halo and thick-disc region. Here,
all three stages (RGB, HB, and AGB stars) attain very large Li
abundance values between $4.0$ and $5.0$ dex, with even a very low
metallicity AGB or an upper RGB  star exhibiting a Li abundance of $6.15$ dex (see below).
Concerning the very metal-poor stars in the region between metallicities
$-4.0$ and $-3.0$, only three RGB stars have been detected and obtained
from the literature. 

\begin{figure}[h]
\centering
\includegraphics[width=\columnwidth]{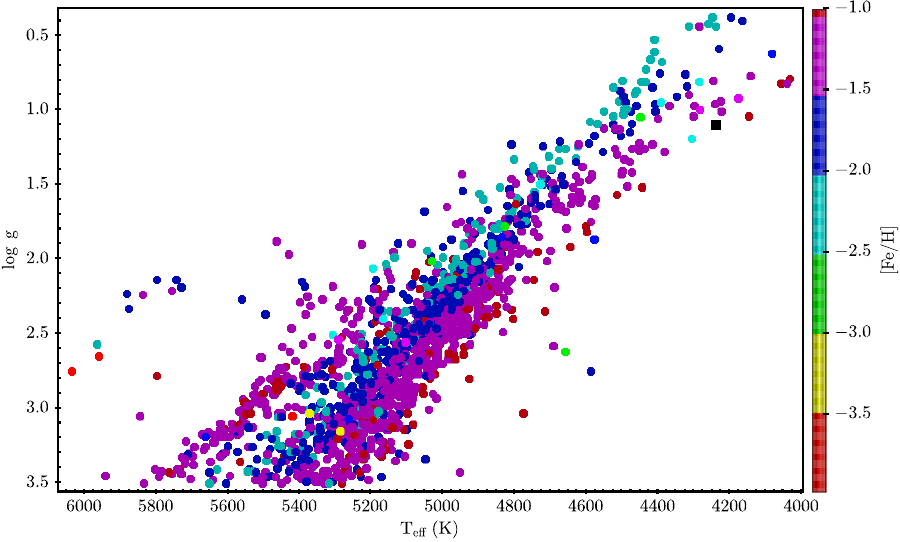}
\vspace{0.3cm}
\includegraphics[width=\columnwidth]{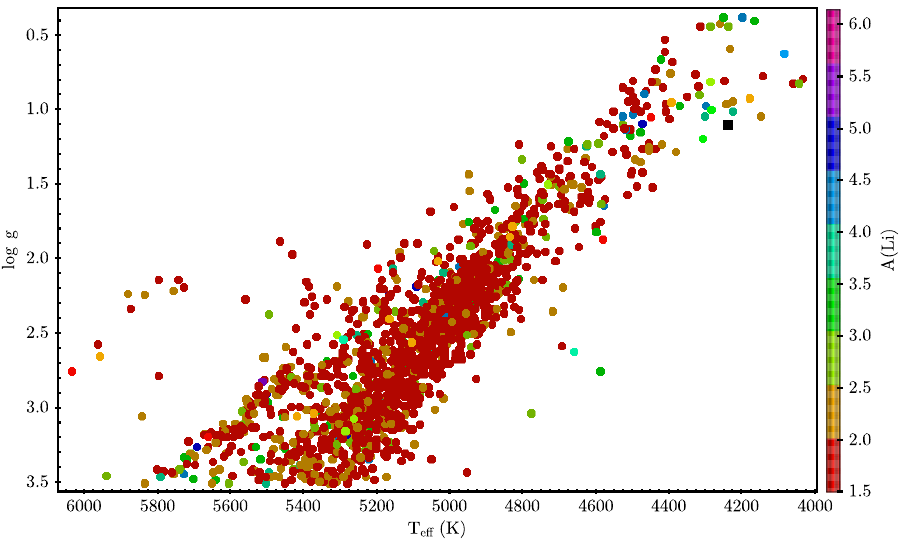}
\caption{
Kiel diagram for Li-rich giant stars, showing $\log g$ as a function of $T_{\text{eff}}$. 
Panel a (top): Metallicity $\mathrm{[Fe/H]}$ shown as the auxiliary axis. 
Panel b (bottom):  Li abundance shown as the auxiliary axis. 
The position of 2MASS J05241392-0336543, the most Li-rich giant star known 
($A(\mathrm{Li}) = 6.15$ dex), is marked as a black square.}
\label{fig:Kiel_diagram}
\end{figure}

Finally, Fig.~\ref{fig:Kiel_diagram}  presents two Kiel diagrams for all Li-rich giants in this
study. These diagrams show $\log g $ as a function of $T_{\text{eff}}$, with auxiliary
axes displaying Li abundance and metallicity. A particularly noteworthy
object is 2MASS J05241392-0336543 with a metallicity of $-2.54$, which
exhibits the highest known Li abundance at $6.15$ dex. \cite{Kowkabany24} noted the challenge of classifying this star as an RGB or AGB
star due to its position in the transition region of these evolutionary
branches. However, given its low $\log g $ value of $1.10$ -typical of AGB
stars, we  can classify it as an AGB star.
 On the other hand, it can also be an upper RGB star. In this respect, it can be
noted that this star presents an important IR excess of 1.60 and that, following our
generalised CF mechanism (see Sect. 4), this giant star is currently producing even more fresh
Li.

The double sequence observed in the lower part of the Kiel diagram is formed
by two distinct stages of the evolutionary path; the red clump (left branch)
and the  RGB (right branch). The stellar distribution in this diagram
is governed by age, metallicity, and, as more recently evidenced, by
kinematics when restricting the sample to objects with the same metallicity
content \citep{Recio_Blanco2024}. In Fig.~\ref{fig:Kiel_diagram}a, it can be seen that the inner edge
of the RGB is populated by the most metal-poor stars. Our
classification into HB, RGB, and AGB is based solely on $\log g$ .

\section{The action of the Cameron-Fowler mechanism} 

The well-known Cameron-Fowler (CF) mechanism \citep{CaFow71} combines several nuclear reactions that result in the formation of \( ^7\text{Li} \). It was originally proposed to explain the observed high Li abundances in AGB giant stars with intermediate masses of approximately four or more solar masses. In such stars, the internal stellar layers where \( ^7\text{Be} \) is formed via \( ^3\text{He} \) are relatively close to the external envelope, facilitating the transport of this element to the surface. Once in the envelope, \( ^7\text{Be} \) is transformed into \( ^7\text{Li} \) through the reaction \( ^7\text{Be} (e^-, \gamma) ^7\text{Li} \). This process, also known as the “hot-bottom process,” has been successfully applied by \cite{SmithLam90}. 

However, in the more numerous low-mass giant stars, with masses below \( 2 M_{\odot} \), even larger quantities of new Li are produced. This necessitated the identification of an alternative mechanism capable of transporting material to deeper internal layers where \( ^7\text{Be} \) is synthesised. \cite{Sackman99} proposed the so-called “cold-bottom process,” in which an ad hoc conveyor-belt circulation mechanism transports envelope material into the deeper layers of the hydrogen-burning zone, where \( ^7\text{Be} \) reaches its peak internal abundance. 

In the literature, it is often assumed that \( ^7\text{Be} \) alone ascends to the surface, where it is converted into \( ^7\text{Li} \). However, simulations of nuclear transport reactions by \cite{Yan18} suggest a more accurate picture of these processes. Their models, designed to replicate Li enrichment in a giant star with an exceptionally high Li abundance of 4.7 dex, indicate that both \( ^7\text{Be} \) and \( ^7\text{Li} \) are transported together, as \( ^7\text{Be} \) is continuously converted into \( ^7\text{Li} \). The rapid upward motion of this Li-rich material is crucial to preventing \( ^7\text{Li} \) destruction by proton reactions during transport. As a result, a new Li-enriched external envelope is formed. 

The nuclear simulations by \cite{Yan18} determined that the required upward transport time for the studied case was 47 years. Based on this timescale, \citetalias{delaReza25} estimated an upper velocity for \( ^7\text{Li} \) transport of approximately \( 470 \,\text{cm/s} \), assuming a stellar atmospheric scale height of \(10~R_{\odot} \). This value is consistent with the framework used by \cite{Eggleton08} to compare their estimated upward velocity of \(2 \,\text{cm/s} \) for carbon transport with that of \( ^7\text{Li} \). However, Yan (private communication) later recalculated the \( ^7\text{Li} \) upward velocity using a more appropriate scale height for his model \citep{Yan18}, finding that the value used in \citetalias{delaReza25} was overestimated by a factor of approximately 20. Nevertheless, the revised average velocity still remains about ten times larger than that of carbon transport. This difference in transport velocities explains the observed anti-correlation between Li abundance and the \( ^{12}\text{C}/^{13}\text{C} \) ratio, as initially noted by \cite{daSilva99}. 

Based on these nuclear processes, \citetalias{delaReza25} proposed a general Li enrichment scenario for low-mass giant stars, in which internal material transport begins in the upper hydrogen-burning zone and extends outward, eventually manifesting as external shells. These external shells, which are observable through significant infrared (IR) excesses, are a consequence of mass loss driven by internal angular momentum evolution. This scenario requires episodic Li enrichment followed by substantial mass-loss events occurring sporadically. The internal mechanism driving this sporadic transport remains unknown, though it has been speculated that magnetic instabilities might play a role. 

A crucial implication of this revised perspective on the CF mechanism is its potential connection to IR excesses and mass loss. In other words, detecting an IR excess in a Li-rich giant star may indicate that the star is currently undergoing Li enrichment. This idea was explored in \citetalias{delaReza25}, which focused on Li-rich giant stars in the thin disc with higher metallicities. In that study, based on 10,535 Li-rich giant stars from the LAMOST-\textit{Gaia} catalogue \citep{gao19}, Li abundances ranged from \( 1.5 \) to \( 4.9 \) dex. Among these, 196 RGB and 421 clump giants exhibited IR excesses, representing 5.8\% of the total Li-rich sample. These findings suggest that these stars are actively enriching their envelopes with Li while simultaneously undergoing episodic mass loss over short timescales of approximately \( 10^4 \) years. Notably, as discussed in \citetalias{delaReza25}, after this brief period of \( ^7\text{Li} \) enrichment, stars retain their Li for a significantly longer timescale of \( 10^7 \) years in the thin disc.

In Table~\ref{tab:tab2}, we present 1703 Li-rich giant stars. Importantly, stars with IR excesses are uniformly distributed in the three evolutionary stages (RGB, HB, and AGB) and within the metallicity range of \( -2.5 \) to \( -1.0 \). These excesses were identified using WISE data by measuring the difference in WISE colours \( W1 - W4 \), where \( W1 \) corresponds to \( 3.4 \,\mu \text{m} \) and \( W4 \) to \( 22 \,\mu \text{m} \). The selection criteria followed those of \cite{Martell21} and \citetalias{delaReza25}, requiring \( W1 - W4 \geq 0.5 \) and excluding the upper limit flux values and the contaminated images. 

\begin{table*}
\caption{Properties of the Li-rich stars.}
\renewcommand{\arraystretch}{1.2}

\centering
\scalebox{0.7}{%
\begin{tabular}{ccccccccccccccc}
\hline\hline
Gaia\tablefootmark{a} & RA\tablefootmark{b} & Dec\tablefootmark{b} &
$T_\mathrm{eff}$\tablefootmark{c} & $\log g$\tablefootmark{d} &
[Fe/H]\tablefootmark{e} & A(Li)\tablefootmark{f} &
[$\alpha$/Fe]\tablefootmark{g} & \( W1 - W4 \)\tablefootmark{h} &
X$_\mathrm{GC}$\tablefootmark{i} & Y$_\mathrm{GC}$\tablefootmark{i} &
Z$_\mathrm{GC}$\tablefootmark{i} & $V_\omega$\tablefootmark{j} &
$V_\phi$\tablefootmark{j} & $V_Z$\tablefootmark{j} \\
\hline
2743690396484004352 & 358.858375 & 5.131389  & 5302 & 3.28 & $-$1.19 & 1.8 &      & 2.46 & 8286.88 & 612.02  & $-$883.31 & 224.42   & $-$41.11 & 15.74 \\
2739765414851064704 & 358.696208 & 2.789472  & 4992 & 2.29 & $-$1.73 & 3.4 &      & 4.38 & 8537.02 & 3110.54 & $-$4838.27 & 84.39   & $-$76.80 & 195.88 \\
2746209480702632960 & 358.643000 & 7.135889  & 5280 & 3.07 & $-$1.21 & 2.2 & 0.26 & 1.97 & 8334.59 & 849.49  & $-$1144.42 & 169.10  & $-$10.76 & 70.66 \\
2867554470161966464 & 358.594000 & 30.456611 & 4933 & 2.24 & $-$1.47 & 1.9 & 0.40 & 3.69 & 9762.05 & 4638.55 & $-$2920.42 & $-$61.76 & $-$3.51 & 12.39 \\
2867650471271000576 & 358.474958 & 30.967194 & 5001 & 2.32 & $-$1.27 & 2.3 &      & 2.54 & 9467.33 & 3755.42 & $-$2316.41 & 84.80   & $-$67.37 & 61.97 \\
.....               & .....       & .....     & .....& .....& .....   & ....& .... &      & .....   & .....   & .....      & .....    & .....    & ..... \\
\hline
\end{tabular}%
}

\tablefoot{
\tablefoottext{a}{\textit{Gaia} DR3 source identifier.}
\tablefoottext{b}{Right ascension and declination in degrees (J2000).}
\tablefoottext{c}{$T_\mathrm{eff}$: effective temperature in K.}
\tablefoottext{d}{$\log g$: surface gravity in dex.}
\tablefoottext{e}{[Fe/H]: metallicity.}
\tablefoottext{f}{A(Li): Li abundance.}
\tablefoottext{g}{[$\alpha$/Fe]: alpha-element enhancement.}
\tablefoottext{h}{\( W1 - W4 \): WISE IR colour index in mag.}
\tablefoottext{i}{X$_\mathrm{GC}$, Y$_\mathrm{GC}$, and Z$_\mathrm{GC}$: positions in pc w.r.t. Galactic Centre.}
\tablefoottext{j}{$V_\omega$, $V_\phi$, $V_Z$: Galactocentric velocity components in km\,s$^{-1}$. This table is shown only in part; the complete version is available online as a Vizier catalogue. }
}
\label{tab:tab2}
\end{table*}

Could the CF mechanism also be active at even lower metallicities (\(-4.0\) to \(-2.5\))? While there are three confirmed Li-rich RGB stars in this metallicity range (see Fig.~\ref{fig:Li_trends}-b), the extreme scarcity of stars in this regime stands in the way of a robust statistical analysis of IR excess frequency and represents a main caveat. However, if the mechanism is indeed  general, it should extend throughout the entire history of the Milky Way. 

\begin{figure}[h]
\centering
\includegraphics[width=\columnwidth]{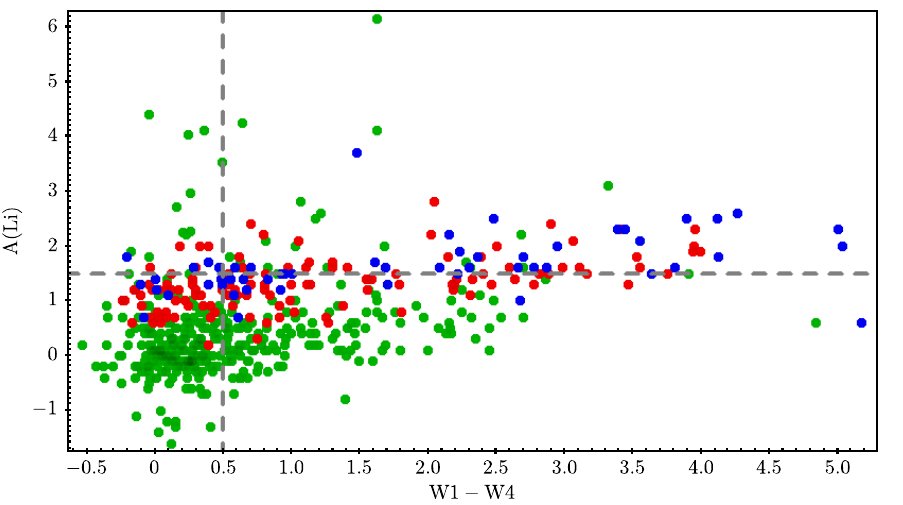}
\caption{Li abundances of giant stars with IR excesses as a function of the WISE colour index  \( W1 - W4 \). Blue points correspond to RGB stars, red to HB stars, and green to AGB stars.}  \label{fig:Li_IR_excess}
\end{figure}

Figure~\ref{fig:Li_IR_excess} presents the IR excesses for the three evolutionary stages (RGB, HB, and AGB), revealing values as high as five magnitudes, similar to those observed in the thin disc \citepalias{delaReza25}. Since these excesses correspond to significant mass-loss episodes, they contribute to Li enrichment in the interstellar medium of the halo/thick disc. 

Concerning this contribution, we have to consider two properties of the mechanisms of Li enrichment. These have been briefly discussed in  \citet{delaReza15} and especially in DLR25. One is a positive contribution in the sense that, due to the proposed universality of Li production, all (at least for the most numerous low-mass giant stars) contribute to the Li enrichment of the ISM. The other is less positive, in the sense that the process of Li enrichment is very short ($\sim 10^{4}$ years or less), even with episodes of strong mass losses on the order of $10^{-7}$--$10^{-6}\,M_{\odot}\,\mathrm{yr}^{-1}$. In view of this, proper calculation of the yield of Li must be estimated and formulated for Galactic chemical models, and this is beyond the scope of our work.

In Fig.~\ref{fig:Li_IR_excess}, we present the Li abundances as a function of the IR excesses for an ensemble of giant stars selected based on the availability of \( W1 \) and \( W4 \) magnitudes with associated errors in the WISE catalogue, as well as their inclusion in the catalogues used in this study. A total of 535 stars were selected, divided into three groups with their respective numbers of objects and colours: AGB (367 green), HB (118 red), and RGB (50 blue). Most AGB stars are located in the first quadrant, exhibiting no IR excess and low Li content. A smaller number of HB and RGB stars show similar behaviour in the first quadrant. The remaining stars in the second quadrant have IR excesses  \( W1 - W4 \) $\ge$ 0.5, representing the true near-IR (NIR) physical excesses considered in this work. As mentioned before in the context of our general CF mechanism of Li enrichment, the presence of a star with an IR excess indicates that the star is undergoing an active Li enrichment process episode.

Figure~\ref{fig:Li_IR_excess} clearly shows the notable presence of three different Li thresholds, each corresponding to the three stages of evolution: for RGB stars (blue) at $\sim$ 1.5~dex, for HB stars (red) at $\sim$ 0.5~dex, and for AGB stars (green) at $\sim$ -0.5~dex. To our knowledge, the presence of three different Li thresholds that appear in low metallicity is unique.   Additionally, there is a general tendency for Li abundance to increase with IR excess, as measured by \( W1 - W4 \). This supports our generalised CF mechanism and suggests that Li enrichment is directly related to the total energy required to transport stellar material from the stellar interior to the exterior \citepalias[see][]{delaReza25}.
 In this case, the energy requirements differ somewhat for each stage, with the AGB stage being the least energetic. 

Accepting these three different Li thresholds and examining Fig.~\ref{fig:Li_IR_excess}, we obtain the following scenario. First, for AGB stars: those located in the second quadrant are considered AGB Li-rich giants currently being enriched with fresh Li. Those in the first quadrant, which show no IR excess, can also be regarded as Li-rich giants, although in these stars the Li enrichment process has already ceased. Notably, in this case there are two super Li-rich AGB stars with Li abundances larger than 4.0 dex. The same scenario also applies to HB and RGB stars.

If our Li-thresholds are correct, the Li-timescales could be different than that of the thin disc which is around $10^{7}$ yrs. However, the knowledge of their values requires more detailed quantitative calculations.

\section{Stellar rotation in the halo and thick disc}

In recent literature, \citet{Mu24} reported the lack of stellar rotational velocity data required to carry out studies in the low-metallicity regime of giant stars. To our knowledge, the present work represents one of the first steps in addressing this issue by utilising two recent large data collections: on the one hand, the LAMA LAMOST catalogue \citep{Li24} for the $v \sin i$ data and, on the other hand, the LAMOST DR9 catalogue \citep{Ding24} for the Li abundance data. In the Galactic thin disc, studies on the rotational properties of Li-rich giant stars and their correlation with Li abundances remain unsatisfactory due to several fundamental unresolved processes. Some studies have approached this problem by employing direct atmospheric observations \citep[e.g.][]{Frasca22} and stellar spot analysis \citep{Kriskovics23}. A general issue has emerged, indicating an apparent lack of correlation between these two parameters. The most comprehensive study on this topic so far is that of \citet{Deepak20}, who utilised a large sample of Li-poor and Li-rich giant stars from the GALAH survey, finding an anti-correlation between Li abundance and $v \sin i$ for three stellar types: Li-poor, Li-rich, and super Li-rich giants (see their figure 7). Regarding the physical mechanisms underlying this open problem concerning the Li versus $v \sin i$ relation, some have been discussed in \citetalias{delaReza25}, but a broader perspective is provided here. At least three key mechanisms are at play: (a) the transfer of total stellar angular momentum (AM); (b) stellar rotation as a driver of activity, and (c) the influence of rotation on the internal reservoir of \( ^3\text{He} \), which serves as the source of \( ^7\text{Li}\).

The first mechanism suggests that the final rotational velocity of the outer stellar envelope depends on how the total stellar AM transfer occurs. Since asteroseismology techniques have revealed that the cores of giant stars rotate slowly \citep[see review by][]{Ae19}, a general picture has emerged in which the stellar envelope rotates roughly ten times slower than the core, as opposed to the previously assumed several orders of magnitude. This new paradigm aligns with the mean rotation of normal Li-poor giant stars, which exhibit $v \sin i$ values below $5\,\mathrm{km\,s^{-1}}$. However, this scenario does not apply to giant stars at low metallicities, which exhibit significantly larger $v \sin i$ values. This discrepancy presents a challenging problem, as different AM transfer mechanisms are required to explain the rapid rotation observed in these stars.

The second key aspect relates rotation to stellar activity. In general, Li-poor giant stars are relatively inactive, exhibiting low mass loss. In contrast, Li-rich giant stars tend to be more active. \citet{Dixon20} showed that a subset of quiet, normal giant stars exhibit significant UV excesses, with this activity being correlated with rotation, as observed in dwarf stars. Consequently, rotation in some giant stars emerges as a driver of stellar activity, manifested in UV emission.
A recent work by \citet{Dixon25} extends this relation from solar metallicities down to -1.5. How do Li-rich giant stars fit into this scenario? \citetalias{delaReza25} showed that among Li-rich and super Li-rich giant stars, Hubble UV emission lines at $\sim2500$ Å exhibit a correlation between line intensity and $v \sin i$. This behaviour is interpreted as a consequence of mass transport from the interior, activating the chromosphere as material migrates outward. This Li-activity relation in giant stars has also been studied using chromospheric He\,I at 10830\,\AA{} \citep{Sneden22, Mallick25}.

The third point, concerning the impact of rotation on \( ^3\text{He} \) abundance, is primarily a topic for internal stellar modelling. The internal \( ^3\text{He} \)-rich zone forms during the main-sequence phase and serves as the source of \( ^7\text{Li}\) through nuclear reactions: $^3He + ( ^4He, \gamma) \rightarrow\  ^7Be $  and $ ^7Be + (e^-, \gamma) \rightarrow\ ^7Li$.

\citet{Eggleton06, Eggleton08}, without considering rotational effects, first explained why \( ^3\text{He} \) production is limited and largely destroyed during the RGB phase, ultimately constraining \( ^7\text{Li}\) overproduction. Later studies \citep{Stancliffe09, Lagarde11} incorporated thermohaline mixing processes to explain \( ^3\text{He} \) destruction in RGB and AGB low-mass stars.

Although stellar rotation in the thin disc typically reaches maximum observed values around $35\,\mathrm{km\,s^{-1}}$ \citep{Reddy05}, significantly higher values are observed at very low metallicities. This could be due to the smaller physical size of stars at low metallicities, allowing for faster rotation. 

Nevertheless, for lower rotational velocities ($v \sin i < 40\,\mathrm{km\,s^{-1}}$), AGB and HB stars can exhibit significantly enhanced Li abundances, reaching values around 5.0 dex. The underlying physical mechanisms
governing the overall distribution of Li abundances and their correlation with $v\sin i$, particularly the existence of this plateau, present a challenging problem in the context of angular momentum transport, as outlined in the preceding discussion. 

\begin{figure}[h]
   \centering
  \includegraphics[width=\columnwidth]{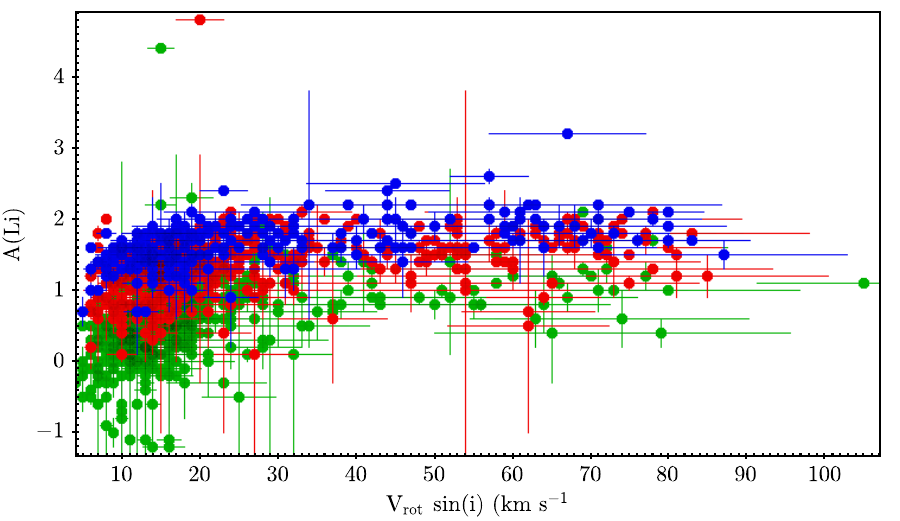}
   \caption{Li abundances of giant stars as a function of stellar rotation velocities. The most prominent feature is a plateau for $v \sin i > 40\,\mathrm{km\,s^{-1}}$, where no super Li-rich RGB (blue) and HB (red) stars with Li abundances exceeding 3.2 dex are found.}
   \label{fig:Li_vs_vsini}
\end{figure}

Figure~\ref{fig:Li_vs_vsini} displays the Li abundance as a function of $v \sin i$,
where a few giant stars with $v \sin i$ below $40\,\mathrm{km\,s^{-1}}$ reach Li abundances typical of very Li-rich giants. In reality, there are many more giants of this type, as illustrated in Fig~\ref{fig:Li_trends}-b. They are not present because the LAMOST catalogue of \cite{Li24} has not measured their $v \sin i$ values. 

 If we examine Figs. \ref{fig:Li_Teff}, \ref{fig:Li_IR_excess}, and \ref{fig:Li_vs_vsini}, considering the values of A(Li) as functions of $T_{\text{eff}}$, \( W1 - W4 \), and 
 $v\sin i$ for the three evolving phases (AGB, HB, and RGB), we can see that in each case, the distributions of A(Li) attain a maximum asymptotic value as their abscissae increase. The only case where the distribution is characterised well by a plateau is the one shown in Fig.~\ref{fig:Li_vs_vsini} for $v\sin i$ values greater than $40\,\mathrm{km\,s^{-1}}$.

  To characterise the plateau, we followed the recommendation of  \citet{Li24}, considering only those values for which the FLAGS assigned to each physical variable are equal to zero. Specifically, we selected stars with both \text{FLAG\_MAD\_VSINI} = 0  and  \text{FLAG\_DELTA\_VSINI} = 0.
  Our sample exhibits a median  of approximately 0.1, with a standard deviation of 0.3. 
  The plateau values corresponding to the three different selections are listed in Table 3. They were derived as the medians of the $v \sin i$  $>$ $40\,\mathrm{km\,s^{-1}}$ distributions. In Fig.~\ref{fig:Li_vs_vsini}, these values are shown using the same colour scheme as that adopted for the metal-poor giant classes.  The statistical values for this case are presented in Table \ref{tab:tab3}. 
  
\begin{table}[ht]
    \centering
    \caption{Statistical properties of the different stellar classes.}
    \begin{tabular}{lccccc}
        \hline\hline
        Class                  & N   & A(Li) median    & A(Li)SD  &     A(Li) MAD \\
        \hline
        \textcolor{green}{AGB} & 41  & 1.1             & 0.30     &  0.2 \\
        \textcolor{red}{HB}    & 87  & 1.6             & 0.30     &  0.2 \\
        \textcolor{blue}{RGB}  & 59  & 1.8             & 0.30     &  0.2 \\
        \hline
    \end{tabular}
    \label{tab:tab3}
\end{table}
 
The ensemble of Fig.~\ref{fig:Li_vs_vsini}  appears to indicate two important conclusions: only
slow giant rotators produce high Li abundances, while rapid and very rapid giant
rotators produce a limited amount of Li. However, a complete confirmation of this
statement requires measurements of the $v \sin i$ values of a large
part of the remaining very Li-rich giants.

Concerning the second effect mentioned above: that is, Li-rich giant stars as stellar activators. The point here is to show in this figure whether Li-rich giant stars, with Li abundances greater than or equal to $ 1.5 $ dex, eventually present indications of activity appropriate for these stars. This can be achieved by means of different known indices such as $ R^{\prime}_{\text{HK}}$, $R^{\prime}_{\text{H$_\alpha$}}$, $R^{\prime}_{\text{NUV}}$, and $R^{\prime}_{\text{FUV}}$, which have been applied recently only to stars belonging to the Galactic thin-disc \citep{Han24}. They could be useful in our case if they were applicable to metallicities lower than $-1.0$. In particular, a comparison between the emission at far-UV (FUV) produced by the external stellar regions and that near-UV (NUV) with photospheric origin is needed. This task remains to be done for a large number of giant stars in the halo and thick disc and could probably provide better insights into the activity of Li-rich giant stars at very low metallicities.

\begin{figure}[h]
\centering

\includegraphics[width=\columnwidth]{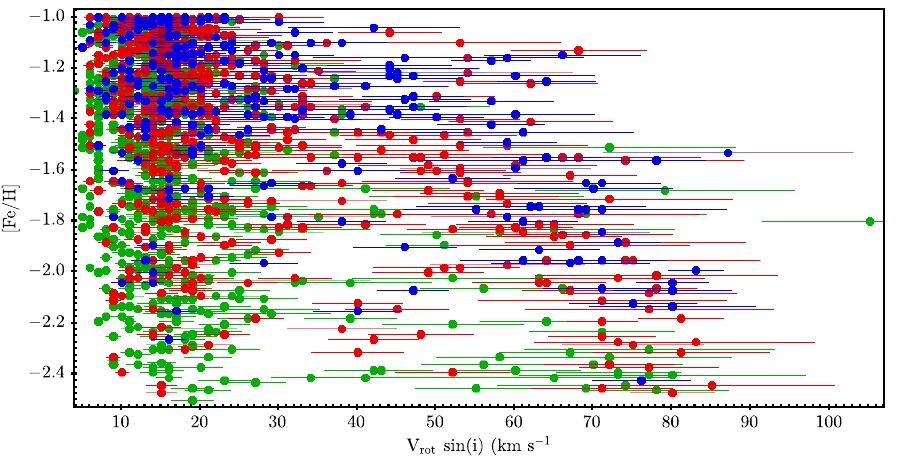}
\caption{Metallicity $\mathrm{[Fe/H]}$ is shown as a function of the rotational velocity. The branch of stars starting at approximately $40\,\mathrm{km\,s^{-1}}$ and showing an increase in velocity up to nearly $90\,\mathrm{km\,s^{-1}}$ as $\mathrm{[Fe/H]}$ decreases from around $-1.0$ to $-2.5$ is composed mainly of Li-rich RGB and HB stars. Colours and symbols are the same as in Fig.~\ref{fig:Toomre}.}
\label{fig:met-rot}
\end{figure}

\begin{table*}
\caption{
Joint table of stellar parameters from LAMOST and \textit{Gaia} catalogues.
}
\renewcommand{\arraystretch}{1.4}  
\resizebox{\textwidth}{!}{%
\begin{tabular}{rrrrrrrrrrrrrrr}
\hline\hline
Gaia\tablefootmark{a} & RA\tablefootmark{b} & Dec\tablefootmark{b} & $T_{\mathrm{eff}}$\tablefootmark{c} & $\log g$\tablefootmark{d} & [Fe/H]\tablefootmark{e} & A(Li)\tablefootmark{f} & $v\sin i$\tablefootmark{g} & $e_{v\sin i}$\tablefootmark{h} & $X_\mathrm{GC}$\tablefootmark{i} & $Y_\mathrm{GC}$\tablefootmark{i} & $Z_\mathrm{GC}$\tablefootmark{i} & $V_\omega$\tablefootmark{j} & $V_\Phi$\tablefootmark{j} & $W_0$\tablefootmark{j} \\
\hline
2875071934240323456 & 359.89 & 33.60 & 4759 & 2.01 & -1.11 & 0.7 & 11 & 0.71 & 9170.80 & 2569.00 & -1462.36 & 46.63 & 14.63 & -45.05 \\
2873798901637520896 & 359.86 & 32.01 & 4912 & 2.22 & -1.36 & 1.4 & 10 & 1.12 & 8961.84 & 2063.16 & -1248.17 & -329.68 & 142.09 & -17.92 \\
2743433832317754880 & 359.79 & 4.61  & 4618 & 1.45 & -1.14 & 0.5 & 13 & 1.58 & 8370.52 & 1044.74 & -1560.16 & 131.56 & 142.56 & -12.39 \\
2739900753564212480 & 359.72 & 2.70  & 4917 & 2.22 & -1.32 & 1.3 & 10 & 0.86 & 8393.49 & 1401.66 & -2228.63 & 251.53 & -98.31 & 177.07 \\
2878564532925873664 & 359.43 & 35.78 & 4580 & 1.16 & -1.76 & 0.9 & 16 & 0.93 & 10845.84 & 6946.74 & -3598.19 & -70.05 & 194.08 & 35.97 \\
...                 & ...    & ...    & ... & ... &... & ... & ... & ... & ... & ... & ... & ... & ... & ... \\
\hline
\end{tabular}%
}
\tablefoot{
\tablefoottext{a}{\textit{Gaia} DR3 source identifier.}
\tablefoottext{b}{Right ascension and declination in degrees (J2000).}
\tablefoottext{c}{$T_\mathrm{eff}$: effective temperature (K).}
\tablefoottext{d}{$\log g$: surface gravity (dex).}
\tablefoottext{e}{[Fe/H]: metallicity (dex).}
\tablefoottext{f}{A(Li): Li abundance (dex).}
\tablefoottext{g}{$v\sin i$: projected rotational velocity (km\,s$^{-1}$).}
\tablefoottext{h}{$e_{v\sin i}$: error in $v\sin i$.}
\tablefoottext{i}{$X_\mathrm{GC}$, $Y_\mathrm{GC}$, $Z_\mathrm{GC}$: Galactocentric positions (pc).}
\tablefoottext{j}{$V_\omega$, $V_\Phi$, $W_0$: Galactocentric velocity components (km\,s$^{-1}$). This table is shown in part; the complete version is available online as a Vizier catalogue.}
}
\label{tab:tab4}
\end{table*}

In the following, we proceed by taking the Li abundance values and the other stellar parameters, such as $T_{\text{eff}}$, $\log\ g$, and [Fe/H] from LAMOST DR9 \citep{Ding24} and the \( v \sin i\) values from LAMA LAMOST \citep{Li24}. By selecting  giant stars according to our criteria based on $\log\ g$, as mentioned in Sect. 2, we obtain a collection of 2,468 Li-rich and Li-poor giant stars presented in Table~\ref{tab:tab4}. Displaying these stars with their [Fe/H] as a function of the \( v \sin i\) values, we plot Fig.~\ref{fig:met-rot}, where, first, for low velocities up to $40\,\mathrm{km\,s^{-1}}$, the velocities do not change as metallicity decreases. Second, for velocities larger than around $40\,\mathrm{km\,s^{-1}}$, stars appear to show an increase in their \( v \sin i\) velocities as metallicity diminishes from $\sim -1.0$ down to $\sim -2.5$.with terminal velocities near $90\,\mathrm{km\,s^{-1}}$. Note that this behaviour corresponds   to Li-rich giant stars.
One tentative interpretation of this behaviour could be that stars diminish in physical size as their metallicity decreases. The reason for this can be found in the formation of the initial Population II stars with very low metallicities. \citet{Dopcke11,Dopck13}, using hydrodynamic simulations, found that under those conditions, dust can be very efficient in cooling the gas and provoking the fragmentation of a cloud. These processes result in the formation of stars with very low masses between $0.01$ and $1$~\(M{\odot}\), which, as expected, can be small-sized stars; however, no stellar radii are provided in this study. Some Galactic models, such as that proposed by \citet{Chiappini04}, even suggest that at extremely low metallicities, stars of these small sizes could rotate with velocities as high as around $700\,\mathrm{km\,s^{-1}}$. In any case, in this work, we are presenting a more realistic observational picture of the stellar rotational scenario for the entire Galaxy. 

A new vision on the explanation
of the high giant rotators and their sizes appears with the recent work by \citet{Mosser2024}, which solves the problem of stellar angular momentum for HB giants by
conserving it. In this way, the stellar core and its envelope spin in the same
manner. For high rotators, this requires small stellar radii. However, all these
properties of high-velocity rotators can be tested by
measuring the stellar radii of these objects.

\section{Discussion and conclusions}

Until now, the stellar evolution of field low-mass Li-rich giant stars at very low metallicities has remained a largely unexplored area of research. We studied the main properties of these sources, which produce fresh $^7\mathrm{Li}$ during their advanced evolutionary stages in the Galactic halo and thick disc. Additionally, we explored, for the first time in this Galactic region, the general behaviour of the stellar rotational velocities of both Li-rich and normal Li-poor giant stars.

For Li studies, the metallicities fall within the range of $-4.0$ to $-1.0$, whereas for the rotational analysis, the range considered is from $-2.5$ to $-1.0$. Two major data sets were used for this purpose: for Li-rich giant stars, the LAMOST DR9 catalogue \citep{Ding24}, and for the $v \sin i$ data of both Li-rich and Li-poor giant stars, the LAMA LAMOST catalogue \citep{Li24}. One challenge in studying the halo and thick disc is the rarity of these stars in the metallicity range of $-4.0$ to $-3.0$, which account for only 0.1\% of the stars in the Galaxy \citep{Yao24}.
 We summarise the main findings of our study below.

\begin{itemize}

\item The distribution of giant stars in the Galactic phase space:\ The Toomre  diagram in Fig.~\ref{fig:Toomre} shows that  most of the stars in the sample belong to the halo and thick disc, with only a few candidates likely to be members of the metal-poor thin disc. 
Based on the data, we obtained the following classifications: halo (74\%), thick disc (22\%), and thin disc (4\%). In Fig.~\ref{fig:Z_Omega}, the spatial distribution shows that AGB stars occupy a larger volume, whereas HB and RGB stars exhibit a more concentrated distribution. This structure simply reflects the intrinsic luminosity gradient associated with the evolutionary stage of the giant stars in our sample (see Fig.~\ref{fig:Li_Teff}). AGB stars, being the most luminous, are visible at greater distances. A potential (yet undetected) influence on Li enrichment could arise from the differing spatial distributions of the three evolutionary stages considered.\\

\item Detection of deficient alpha-element giant stars in the halo and thick disc:\ As observed (Fig.~\ref{fig:alfa_Fe}) and as expected for the low metallicities characteristic of the halo and thick disc, the $[\alpha/\text{Fe}]$ ratio as a function of [Fe/H] is positive and generally exceeds 0.1. However, a few giant stars display negative ratios, indicating a deficiency in these elements. These relatively rare objects, found at very low metallicities, are particularly noteworthy as they suggest early enrichment by Type Ia supernovae as well as by core-collapse supernovae that are deficient in $\alpha$-elements \citep{Jeena2024}. The presence of such giants across the metallicity range from $-2.5$ to $-1.0$ provides a valuable opportunity to investigate the rate and nature of supernova contamination during the evolution of the halo and thick disc.\\

 \item Relation to temperature:\ In Fig.~\ref{fig:Li_Teff}, we can see that all the Li abundances appearing in this work exhibit a
clear relation with the important parameter, $T_{\text{eff}}$, which clearly separates the three evolutionary
stages: AGB, HB, and AGB.\\

\item The stellar evolutionary stages of Li-rich giant stars:\ At extremely low metallicities, these stars resemble fossils of the earliest Population II stars. Even under such conditions, only three  Li-rich RGB stars have been identified within the metallicity range of $-4.0$ to $-3.0$ (see Fig.~\ref{fig:Li_trends}-b). It is noteworthy that these `fossil' RGB Li-rich stars exhibit Li abundances ranging between 2 and 3 dex, indicating that they have undergone Li enrichment through an as-yet unidentified mechanism.
We propose that this mechanism corresponds to the well-known Cameron-Fowler nuclear process for \( ^7Li \)
production and that it was already active in the early history of the Galaxy. It is remarkable that the three
evolutionary stages (RGB, HB, and AGB) participate in Li
production in the halo and thick disc; furthermore, this is also the case in the thin-disc, whereby  Li production is maintained by giant stars throughout the entire
Galaxy. This is valid even if HB and AGB stars    are only  detected  at metallicities around $-2.7$ (see Fig.~\ref{fig:Li_trends}-b), as these stars exhibit the highest observed Li abundances, spanning to a maximum of 6.15 dex. These evolutionary stages play a  role in Li production via episodic, substantial mass loss, enriching the interstellar medium in the halo/thick disc, just as giant stars do in the thin disc.\\

\item The general validity of the Cameron-Fowler mechanism for $^7\mathrm{Li}$ formation across the Galaxy:\ In \citetalias{delaReza25}, a general framework for the CF mechanism in the thin disc is presented, detailing the various steps of Li enrichment in giant stars as a result of stellar angular momentum evolution. This includes the transport of $^7\mathrm{Be}$ and $^7\mathrm{Li}$ from the internal upper H-burning layer to the stellar surface, forming circumstellar shells with short-lived episodes of intense mass loss, observable as IR excess. Given that IR excesses have been detected throughout the metallicity range $-2.5$ to $-1.0$, we argue that the CF mechanism is active in this interval.
In the extreme  low metallicity range of $-4.0$ to $-3.0$, only three RGB Li-rich giants have been detected due to the extreme scarcity of  Galactic stars in this metallicity region. However, since these few stars exhibit Li enrichment, we infer that the CF mechanism is also at work there. If confirmed, this would imply that the CF mechanism operates universally across the entire Galaxy.\\

 \item  Detection of three Li-thresholds:\ By means of the general CF mechanism and 535 giant stars with good quality WISE data to measure their IR excesses, we found the existence of three Li thresholds corresponding to the three giant evolving phases: for RGB at 1.5~dex, for HB at $\sim 0.5$~dex, and for AGB at $\sim -0.5$~dex. The Li formation processes from these thresholds is considered to be independent for each phase. Additionally, they show a tendency to produce more Li as the IR excesses increase (see Fig.~\ref{fig:Li_IR_excess}), indicating a progressive energy requirement for the complete evolution of stellar angular momentum to transport mass from the interior to the external parts of the star, forming fresh $^7\mathrm{Li}$  and creating circumstellar shells. These energy requirements appear to progress from AGB to RGB stages. If all of these findings are indeed correct, these triple thresholds would completely change the notion of Li richness in the low metallicity regime. While RGB stars surprisingly maintain the well-known 1.5~dex Li threshold, this is not the case for the HB and AGB stages. In particular, AGB stars with such tiny quantities of Li are characterised as Li-rich, with the majority of them fitting this description.\\

\item Activity of Li-rich giant stars:\ While normal Li-poor giant stars are generally quiescent, Li-rich giants are active, with their activity potentially driven   especially by rotation.
Unfortunately, there is an almost complete absence of published appropriate
spectral activity indicators in the low-metallicity regime considered in this
work. This impedes any progress in the study of the relation between Li and
activity.\\

 \item The presence of a plateau for larger rotation velocities:\ The behaviour of Li abundances as a function of the rotational velocities (Fig.~\ref{fig:Li_vs_vsini}) presents two different scenarios depending on whether $v \sin i$ is smaller or larger than $40\,\mathrm{km\,s^{-1}}$, which appears to be a critical velocity. For smaller values, RGB, HB, and AGB stars exhibit the entire range of Li abundances from $\sim -2.0$ up to $5.0$~dex. This indicates that only low-velocity giant rotators are capable of producing Li-rich and super Li-rich giants. On the contrary, large rotators form a plateau that spans a wide range, encompassing Li abundances from $\sim 0.5$ to $\sim 2.2$~dex, with no stars exhibiting high Li abundances above $3.2$~dex.
Regarding very rapid rotators, we propose that this behaviour could be explained by a progressive diminishment of the star’s size, allowing them to spin at an increasingly faster rate. Theoretical simulations of the formation of very low-mass stars at very low metallicities support this scenario. This situation is also supported by a recent solution regarding the transfer of stellar angular momentum for HB giants \citep{Mosser2024}, in which large rotators require smaller stellar radii. Nonetheless, all the mentioned properties of high-velocity rotators can be tested by measuring the stellar radii of these objects. The conclusion that only low rotators can produce very Li-rich giants, however, would require studies involving a greater number of observed values of $v \sin i$ than the sample presented here.
\\

\item The presence of an increase in rotation velocities of Li-rich giant stars with a decrease in metallicities: When the metallicities of giant stars are represented as a function of the rotational velocities 
(Fig.~\ref{fig:met-rot}), again two different scenarios appear for values smaller or larger than the critical $40\,\mathrm{km\,s^{-1}}$ velocity. For slower giant rotators, there is no change in their velocities as metallicity decreases from $\sim -1.0$ to $\sim -2.5$. In contrast, for faster rotators, a mixed group of only RGB and HB stars appears to increase their velocities as metallicity decreases by the same interval. AGB stars also decrease in the same way. The same arguments related to the sizes of the stars, as mentioned before, can be applied here. The necessity to increase the lifetimes of the \( ^3\text{He}\) layers, considering stellar rotations:\ Early studies by \citet{Eggleton06, Eggleton08} explain how the \( ^3\text{He} \) reservoir, which is used to create   \( ^7\text{Li} \), is destroyed during the RGB phase, thereby preventing an overproduction of  \( ^7\text{Li}  \). Nevertheless, these studies do not account for rotation in their models.\\ 

In this work, we present a large sample of RGB and HB Li-rich stars with rotation velocities of up to $\sim $170\,$\mathrm{km\,s^{-1}}$ (most of them below $\sim $90\,$\mathrm{km\,s^{-1}}$). Our findings indicate that future internal stellar models considering those stellar rotations must extend the lifetimes of $^3$He zones where \( ^7\text{Li}  \) is created during the advanced stages of evolution.

\end{itemize}

\section*{Data availability}
 Tables~\ref{tab:tab2} and \ref{tab:tab4} are only available at the CDS via anonymous ftp to \texttt{cdsarc.u-strasbg.fr} (130.79.128.5) or via \url{http://cdsweb.u-strasbg.fr/cgi-bin/qcat?J/A+A/}.

\begin{acknowledgements}
We thank the referee for a detailed review and for the insightful comments and suggestions provided. RdlR wishes to thank H.L. Yan for a calculation included in this paper, and Naomi Snoek and Pacha Guzman Deheza for fruitful conversations.
This research has used the Spanish Virtual Observatory (\url{https://svo.cab.inta-csic.es}) project funded by MCIN/AEI through grant PID2023-146210NB-I00. FLA also thanks the technical support provided by A. Parras (CAB) and the encouraged support offered by R.M. Castro-Sánchez.
This work has been partially funded by the Spanish MCIN/AEI grant PID2022-136640NB-C21.
EJA acknowledges financial support from the Severo Ochoa grant CEX2021-001131-S funded by MCIN/AEI/ 10.13039/501100011033. EJA  has made continuous use of TopCat \citep{TopCat2005}. This work presents results from the European Space Agency (ESA) space mission's \textit{Gaia} data. \textit{Gaia} data are being processed by the \textit{Gaia} Data Processing and Analysis Consortium (DPAC). Funding for the DPAC is provided by national institutions,  particular those participating in the \textit{Gaia} MultiLateral Agreement (MLA). The \textit{Gaia} mission website is \url{https://www.cosmos.esa.int/gaia}. The \textit{Gaia} archive website is \url{https://archives.esac.esa.int/gaia}.

\end{acknowledgements}

\bibliographystyle{aa}
\bibliography{Biblio}

\end{document}